\documentclass[aps,prd,preprint,tightenlines,groupedaddress,nofootinbib,showpacs,byrevtex]{revtex4}
\usepackage{amssymb,latexsym}
\usepackage{amsmath,amsbsy}
\usepackage{epsfig,bm}
\usepackage{graphicx,comment}
\DeclareMathOperator{\tr}{tr}

\begin{document}
\def\a{{\alpha}}
\def\b{{\beta}}
\def\d{{\delta}}
\def\D{{\Delta}}
\def\e{{\varepsilon}}
\def\g{{\gamma}}
\def\G{{\Gamma}}
\def\k{{\kappa}}
\def\l{{\lambda}}
\def\L{{\Lambda}}
\def\m{{\mu}}
\def\n{{\nu}}
\def\o{{\omega}}
\def\O{{\Omega}}
\def\S{{\Sigma}}
\def\s{{\sigma}}
\def\th{{\theta}}

\def\ol#1{{\overline{#1}}}

\def\Dslash{D\hskip-0.65em /}
\def\Dtslash{\tilde{D} \hskip-0.65em /}

\def\CPT{{$\chi$PT}}
\def\QCPT{{Q$\chi$PT}}
\def\PQCPT{{PQ$\chi$PT}}
\def\tr{\text{tr}}
\def\str{\text{str}}
\def\diag{\text{diag}}
\def\order{{\mathcal O}}

\def\cC{{\mathcal C}}
\def\cB{{\mathcal B}}
\def\cT{{\mathcal T}}
\def\cQ{{\mathcal Q}}
\def\cL{{\mathcal L}}
\def\cO{{\mathcal O}}
\def\cA{{\mathcal A}}
\def\cQ{{\mathcal Q}}
\def\cR{{\mathcal R}}
\def\cH{{\mathcal H}}
\def\cW{{\mathcal W}}
\def\cM{{\mathcal M}}
\def\cD{{\mathcal D}}
\def\cN{{\mathcal N}}
\def\cP{{\mathcal P}}

\def\Qt{{\tilde{Q}}}
\def\Dt{{\tilde{D}}}
\def\St{{\tilde{\Sigma}}}
\def\cBt{{\tilde{\mathcal{B}}}}
\def\cDt{{\tilde{\mathcal{D}}}}
\def\cTt{{\tilde{\mathcal{T}}}}
\def\cMt{{\tilde{\mathcal{M}}}}
\def\At{{\tilde{A}}}
\def\cNt{{\tilde{\mathcal{N}}}}
\def\cOt{{\tilde{\mathcal{O}}}}
\def\cPt{{\tilde{\mathcal{P}}}}

\def\eqref#1{{(\ref{#1})}}

 
\title{Flavor Twisted Boundary Conditions and the Nucleon Axial Current}
\author{ Brian C.~Tiburzi}
\email[]{bctiburz@phy.duke.edu}
\affiliation{Department of Physics\\
Duke University\\
P.O.~Box 90305\\
Durham, NC 27708-0305}

\date{\today}


\begin{abstract}
With twisted boundary conditions on the quark fields, we study 
nucleon matrix elements of the axial current utilizing 
twisted heavy baryon chiral perturbation theory. One can 
explore the momentum transfer dependence of the axial form factors 
more easily than by using ordinary lattice quantized momenta alone. 
As examples, we derive expressions for the nucleon axial radius
and pseudoscalar form factor.  
\end{abstract}

\maketitle

\section{Introduction}

Lattice QCD calculations will provide first principles determination of low-energy hadronic properties. 
This low-energy region of QCD is the regime of strong interactions, where perturbation theory fails and the
quark and gluon degrees of freedom are confined into color neutral hadronic states. Advances in lattice QCD have 
largely come from exploiting the features of numerically approximating the theory on a discrete lattice in a finite volume. 
As such, there is freedom to manipulate lattice quantities, so that the QCD answer is recovered in the continuum and infinite volume limits.

A mutable feature of finite volume simulations is the choice of boundary conditions satisfied by the fields. 
Periodic boundary conditions are usually assumed as a matter of convenience, as they lead to 
fields and hence observables which are single valued. Observables are determined, however, from $S$-matrix 
elements and hence only the action need be single valued for physical quantities to be well defined. 
The generator of any symmetry of the action can also be used to specify boundary conditions. In particular symmetries involving flavor
generators lead to so-called twisted boundary conditions on the quark fields. 
The appearance of twisted boundary conditions is not new. They have been considered in various contexts over the years %
\cite{Gross:1982at,Roberge:1986mm,Wiese:1991ku,Luscher:1996sc,Bucarelli:1998mu,Guagnelli:2003hw,Kiskis:2002gr,Kiskis:2003rd,Kim:2002np,Kim:2003xt}.
Recently there has been renewed interest in utilizing twisted boundary conditions in lattice QCD simulations %
\cite{Bedaque:2004kc,deDivitiis:2004kq,deDivitiis:2004rf,Sachrajda:2004mi,Bedaque:2004ax}.
This interest stems from the realization that the restriction to lattice quantized momenta can be bypassed
by employing twisted quark fields.

In this work, we consider the nucleon axial transition form factors. On a given lattice, 
the available momentum transfer is quantized, and on current lattices one has integer multiples of
about a few hundred MeV.
While chiral perturbation theory (\CPT) predicts the momentum transfer dependence of 
these form factors, both the quark masses and lattice momenta must be brought down so that 
one enters the effective theory's range of applicability. Another difficulty encountered with a coarse sampling 
of the momentum transfer on a fixed lattice 
is the determination of radii and moments which are only accessible in the near-forward limit.\footnote{%
To sample smaller values of the momentum transfer, one could work with lattices that are much longer in only  
one spatial direction. With such lattices, however, working with sufficiently small lattice spacings and quark masses 
is well beyond the reach of current simulations.
}  
These issues can be circumvented with twisted quarks.

Before specializing to the case of the nucleon axial matrix elements below,
we begin with a more general observation about single particle matrix elements of flavor-changing operators.\footnote{%
A similar observation can also be made for particle to vacuum transition matrix elements. 
For a generic pseudoscalar $H$, we have schematically $\langle 0 | \bm{\cA} | H(\bm{P}) \rangle = i f_H \bm{P}$, 
where $\bm{\cA}$ is the spatial part of the axial-vector current. With twisted boundary conditions and for a flavor non-diagonal state $H$, 
one can access the decay constant $f_H$ at zero lattice momentum. While one could have easily deduced the decay constant by 
working in the rest frame and using the time-component of the current, 
the above observation provides a novel check on using twisted quarks. %
}
These matrix elements have the form $\langle H' (\bm{P'}) | \cO | H (\bm{P}) \rangle$.
Here the operator $\cO$ changes the flavor composition of the hadron $H$ into that of $H'$, and 
we assume at least one of the states is not flavor neutral.
The above matrix element can be decomposed into various form factors which depend on the momentum transfer.
Let the size of the lattice be $L$ in each of the spatial directions.  
On the lattice, the momentum transfer $\bm{q} = \bm{P'} - \bm{P}$ is quantized $\bm{q} = 2 \pi \bm{n} / L$, 
for $\bm{n} \in \mathbb{Z}^3$, when one imposes periodic boundary conditions on the quark fields.
To smoothly access the momentum transfer dependence of the transition form factors, one is forced to vary the lattice size 
and hence generate new gauge configurations.\footnote{%
In principle, one could use the dynamically generated quark mass dependence of the initial and final state hadron masses 
to explore the momentum transfer dependence of such form factors, assuming that these hadrons are non-degenerate. 
This does not give one direct control over the momentum transfer probed, moreover, the 
procedure is complicated by the additional need for chiral extrapolation.
}
Imposing twisted boundary conditions on the valence quark fields, 
such that different flavors are independently rotated in phase at the boundary, 
one finds an induced momentum transfer $\bm{q} = 2 \pi \bm{n} / L + \d \bm{\th} / L$, 
where $\d \bm{\th}$ is the difference in twist angles of the flavors changed. 
Thus with twisted boundary conditions on the valence quarks, the momentum transfer can be continuously varied
and one can access off diagonal matrix elements, even at zero lattice momentum transfer $\bm{n} = \bm{0}$. 
For such matrix elements, one can work in the near forward limit without utilizing a box with one very long side, and without 
performing a momentum extrapolation.

In the following section (Sec.~\ref{s:obs}), we focus on the nucleon axial matrix elements. 
Here we extend our momentum transfer observation above to the case of the neutron-proton 
axial correlation function.  Next in Sec.~\ref{s:tBXPT}, we develop partially twisted baryon
\CPT\ for lattice calculations in which the valence (and ghost) quarks satisfy twisted boundary conditions,
while the sea quarks remain periodic.  
This effective theory can be used at finite volume to ascertain the corrections to nucleon matrix elements due to twisting. 
As an application, we calculate the axial radius and pseudoscalar form factor of the nucleon in Sec.~\ref{s:ax},
and summarize our findings in Sec.~\ref{summy}.

\section{Nucleon Axial Transition} \label{s:obs}

Let us now focus specifically on the nucleon axial matrix elements and make our introductory 
observation more concrete. For simplicity, here and below we work in 
the isospin limit, and consider the neutron and proton to be degenerate. 
We also frame our discussion using continuum operators. 
The relevant flavor-changing
operator for the nucleon axial transition 
is $\cO_\mu(x) = \ol u(x) \gamma_\mu \gamma_5 d(x)$. The Euclidean time, axial correlation function has the form
\begin{equation}
\langle \cP {}_{\bm{0}}(t) \cO_\mu(t') \ol \cN {}_{\bm{P}}(0) \rangle 
= 
\sum_{\bm{x}, \bm{x'}} 
e^{-i \bm{P}\cdot \bm{x'}}
\langle 0 | \cP(\bm{x},t) \cO_\mu (\bm{x'},t') \ol \cN (\bm{0},0) | 0 \rangle
,\end{equation}
where the sums are over all lattice sites labeled by $\bm{x}$, while $\cN(\bm{x},t)$ and $\cP(\bm{x},t)$ 
are interpolating fields for the neutron and proton, respectively. On the left-hand side of the 
equation, the subscripts on the interpolating fields denote the spatial momentum. 
In writing this correlation function, we have assumed that $t > t' > 0$, and for simplicity have chosen 
the final state to be at rest. As is standard, 
the ground state nucleons will be filtered out in the limit of large Euclidean time separation, $t \gg t' \gg 0$,
and the nucleon axial matrix element can be isolated by taking a ratio of the three-point and two-point functions 
weighted with the appropriate kinematic factor.

Due to the periodicity of the fields, the initial-state momentum is quantized 
$\bm{P} = 2 \pi \bm{n} / L$, and the available momentum transfer probed on a given set of gauge configurations is hence
coarsely grained for current lattice sizes. To circumvent this restriction, 
we replace the interpolating fields built from periodic quark fields with new interpolating fields
formed from twisted quark fields. We shall use the same notation for these fields used above: $\cN(\bm{x},t)$ and $\cP(\bm{x},t)$. 
The precise form of the twisting will be spelled out later in Sec.~\ref{s:tBXPT}. 
It suffices to note that each valence quark flavor independently suffers a discontinuous change of phase at the lattice boundary,
and we denote $\bm{\th}^u$ and $\bm{\th}^d$ as the twisting angles for $u$ and $d$ quarks, respectively.

To implement the twisted boundary conditions, one uses modified fields that are periodic
but coupled to a $U(1)$ gauge field.  For the quarks, one calculates their propagators
in this background gauge field~\cite{deDivitiis:2004kq}, where each flavor's twisting angle 
acts as an induced charge.  For the baryons, we can formally separate off the kinematic effects of twisting by defining 
the periodic interpolating fields
\begin{eqnarray}
\tilde{\cN}(\bm{x},t) &=& e^{-i \bm{B}_{\cN} \cdot \bm{x}} \cN(\bm{x},t), \notag \\ 
\tilde{\cP}(\bm{x},t) &=& e^{-i \bm{B}_{\cP} \cdot \bm{x}} \cP(\bm{x},t),
\end{eqnarray}
where $\bm{B}_\cN = ( \bm{\th}^u + 2 \bm{\th}^d ) / L$ and  $\bm{B}_\cP = ( 2 \bm{\th}^u + \bm{\th}^d ) / L$
are the effective $U(1)$ charges of the neutron and proton, respectively.
Using these periodic fields, one then calculates the various Wick contractions in terms of modified
quark propagators.

In order to determine the axial correlation function, one must
perform the Wick contractions when the axial operator is inserted\footnote{%
Notice there are no self-contractions of the operator $\cO_\mu(x)$.}. 
To do so, one inserts the similarly modified axial-vector operator $\cOt_\mu (x)$ 
which is defined in terms of periodic quark fields with effective $U(1)$ charges. 
This operator is related to the twisted quark axial-vector operator via 
\begin{equation}
\cOt_\mu (x) = e^{-i (\bm{\th}^{d} - \bm{\th}^{u}) \cdot \bm{x} /L} \cO_\mu(x)
.\end{equation}
The calculation of Wick contractions then proceeds without any extraneous position-dependent phase factors. 
To obtain the correlation function using twisted fields on the lattice, one hence determines 
\begin{eqnarray}
&& \sum_{\bm{x}, \bm{x'}} 
e^{-i \bm{P}\cdot \bm{x'}} 
\langle 0 | \cPt(\bm{x},t) \cOt_\mu (\bm{x'},t') \ol \cNt (\bm{0},0) | 0 \rangle
\notag \\ 
&& \phantom{space} = 
\sum_{\bm{x}, \bm{x'}} 
e^{-i \bm{B}_\cP \cdot (\bm{x} - \bm{x'})}
e^{-i (\bm{P} + \bm{B}_\cN) \cdot \bm{x'}}
\langle 0 | \cP(\bm{x},t) \cO_\mu (\bm{x'},t') \ol \cN (\bm{0},0) | 0 \rangle
\notag \\
&& \phantom{space} =
\langle \cP {}_{\bm{B}_\cP}(t) \cO_\mu(t') \ol \cN {}_{\bm{P} + \bm{B}_{\cN}}(0) \rangle 
,\end{eqnarray}
where in the subsequent steps we have rewritten the lattice correlation function to expose that the 
initial and final states have been boosted.
The momentum transfered to the final state is effectively $\bm{q} = (\bm{\th}^u - \bm{\th}^d - 2 \pi \bm{n}) / L $ 
and remains non-zero even at zero lattice momentum, i.e.~when  $\bm{n} = \bm{0}$.

While the momentum transfer in such matrix elements can varied continuously, 
twisting produces long-range flavor symmetry breaking interactions that modify
the physics we seek to explore.  
On the lattice, twisting thus introduces modified finite volume corrections that can be determined using chiral effective 
theories~\cite{Sachrajda:2004mi}. 
To this end, we develop partially twisted heavy baryon chiral perturbation theory and apply it to the nucleon axial matrix elements. 
We ultimately address two simple examples that do not have sizable finite volume corrections.

\section{Partially twisted baryon chiral perturbation theory} \label{s:tBXPT}

To address the consequences of twisting in lattice calculations of baryon properties, we construct the underlying 
effective theory in the baryon sector. First we detail the partially twisted boundary conditions
employed and then proceed to include these effects in heavy baryon \CPT.
The quark part of the partially quenched QCD Lagrangian is
\begin{equation}
\mathfrak{L} = \sum_{j,k=1}^6 \ol{Q}{}^{\hskip 0.2em j} \left(
  i\Dslash - m_Q \right)_j^{\hskip 0.3em k} Q_k
.\label{eq:pqqcdlag}
\end{equation}
The six quark fields transform in the fundamental representation of the
graded $SU(4|2)$ group and appear in the vector 
$Q^{\text{T}} = (u, d, j, l, \tilde{u}, \tilde{d})$.
In addition to the $u$ and $d$ quarks, we have added ghost quarks
$\tilde{u}$ and $\tilde{d}$, which cancel the closed valence loops, and two 
sea quarks $j$ and $l$. 
In the isospin limit, the quark mass matrix of $SU(4|2)$ reads
$m_Q = \diag(m_u, m_u, m_j, m_j, m_u, m_u)$,
so that QCD is recovered in the limit $m_j \rightarrow m_u$.
We require that the quark fields satisfy twisted boundary conditions, namely
\begin{equation}
Q(x + L \hat{\bm{e}}_r) = \exp \left( i \theta^a_r \, \ol T {}^a \right) Q(x)  
,\end{equation}
where $\hat{\bm{e}}_r$ is a unit vector in the $r^{\text{th}}$ spatial direction and 
the block diagonal form of the supermatrices $\ol T {}^a$ is
\begin{equation} \label{eq:qtwist}
\ol T {}^a 
= 
\diag 
\left( T^a, 0, T^a \right)
.\end{equation} 
Here $T^a$ are the elements of the $U(2)$ algebra. 
In the isospin limit, any generator can be chosen for the twists, 
although we choose to preserve electric charge conservation and 
accordingly restrict $T^a$ to the Cartan subalgebra. 
Notice in Eq.~\eqref{eq:qtwist} the sea quarks remain periodic at
the boundary. Consequently the twist angles can be changed without necessitating
the generation of new gauge configurations and the fermionic determinant, which
arises solely from the sea sector, is not affected by the twisting.

Redefining the quark fields as $\Qt(x) = V^\dagger(x) Q(x)$, where $V(x) = \exp ( i \bm{\th}^a \cdot  \bm{x} \, \ol T {}^a / L )$, we
can write the partially quenched QCD Lagrangian as
\begin{equation}
\cL = \sum_{j,k=1}^6 \ol{\Qt}{}^{\hskip 0.2em j} \left(
  i \Dtslash - m_Q \right)_j^{\hskip 0.3em k} \Qt_k
,\end{equation}
where all $\Qt$ fields satisfy periodic boundary conditions, and the effect of twisting has
the form of a gauge field:  $\Dt_\mu = D_\mu + i B_\mu$, where $B_\mu = (0, \bm{\th}^a \, \ol T {}^a / L)$. 
It will be easier to treat the twisting in the flavor basis of the valence and ghost sectors rather than in the generator basis, 
thus we write
$\bm{\th}^a \, \ol T {}^a = \diag (\bm{\th}^u, \bm{\th}^d, \bm{0}, \bm{0}, \bm{\th}^u, \bm{\th}^d  )$, 
and similarly for $B_\mu = \diag (B^u_\mu, B^d_\mu, 0, 0, B^u_\mu, B^d_\mu )$.

In the meson sector of partially quenched \CPT\ (\PQCPT)~\cite{Bernard:1994sv,Sharpe:1997by,Golterman:1998st,Sharpe:2000bc,Sharpe:2001fh}, 
the coset field $\Sigma$, which satisfies twisted boundary conditions, can be traded in
for the field $\St$ defined by
$\St (x) = V^\dagger(x) \S (x) V(x)$,
which is periodic at the boundary \cite{Sachrajda:2004mi}. 
In terms of this field, the Lagrangian of \PQCPT\ appears as
\begin{equation}
\cL = 
\frac{f^2}{8} \str \left( \Dt^\mu \St \Dt_\mu \St^\dagger \right)
+ 
\l \, \str \left( m_Q^\dagger \St + \St^\dagger m_Q \right)
.\end{equation}
The action of the covariant derivative $\Dt^\mu$ is specified by
$\Dt^\mu \St = \partial^\mu \St + i [ B^\mu, \St ]$.

To include baryons into \PQCPT, one uses rank three flavor tensors \cite{Labrenz:1996jy}.
The spin-$\frac{1}{2}$ baryons are described by the $\bm{70}$-dimensional supermultiplet $\cB^{ijk}$, 
while the spin-$\frac{3}{2}$ baryons are described by the $\bm{44}$-dimensional supermultiplet $\cT_\mu^{ijk}$ %
\cite{Beane:2002vq}. 
The baryon flavor tensors are, however, twisted at the boundary of the lattice. 
In the $r^{\text{th}}$ spatial direction, both tensors satisfy boundary conditions of the form
\begin{eqnarray}
\cB_{ijk} (x + \hat{\bm{e}_r} L ) 
&=&  (-)^{\eta_{i'} (\eta_j + \eta_{j'}) + (\eta_{i'} + \eta_{j'} )( \eta_k + \eta_{k'})}
\left( e^{i \theta^a_r \, \ol T {}^a }\right)_{ii'}  
\left( e^{i \theta^a_r \, \ol T {}^a }\right)_{jj'}  
\left( e^{i \theta^a_r \, \ol T {}^a }\right)_{kk'} 
\cB^{i'j'k'}(x) \notag \\
&=&
\left( e^{i \theta^a_r \, \ol T {}^a }\right)_{ii}  
\left( e^{i \theta^a_r \, \ol T {}^a }\right)_{jj}  
\left( e^{i \theta^a_r \, \ol T {}^a }\right)_{kk} 
\cB_{ijk}(x)
,\end{eqnarray}
where the $\eta$'s are the grading factors of $SU(4|2)$.
In the last line, we used the diagonality of the Cartan generators.  
Thus we define new tensors $\cBt^{ijk}$ and $\cTt_\mu^{ijk}$ both having the form
\begin{equation}
\cBt_{ijk}(x) = V^\dagger_{ii}(x) V^\dagger_{jj}(x) V^\dagger_{kk}(x) \cB_{ijk}(x)
.\end{equation}
These baryon fields satisfy periodic boundary conditions and their free Lagrangian has the form
\begin{eqnarray} \label{eqn:L}
  \cL
  &=&
  i\left(\ol\cBt v\cdot \cDt \cBt \right)
  +2\a_M^{(PQ)} \left(\ol\cBt \cBt \cMt_+ \right)
  +2\b_M^{(PQ)} \left(\ol\cBt \cMt_+\cBt \right)
  +2\sigma_M^{(PQ)} \left(\ol\cBt \cBt \right) \str\left( \cMt_+\right)
                              \nonumber \\
  &&
  -i\left(\ol\cTt_\mu v\cdot \cDt\cTt^\mu\right)
  +\D\left(\ol\cTt_\mu\cTt^\mu\right)
  +2\g_M^{(PQ)}\left(\ol\cTt_\mu \cMt_+ \cTt^\mu \right)
  -2\ol\sigma_M^{(PQ)}\left(\ol\cTt_\mu\cTt_\mu\right)\str\left( \cMt_+ \right) 
,\notag \\ \end{eqnarray}
where the mass operator is defined by 
$\cMt_+ = \frac{1}{2}\left(\tilde{\xi}^\dagger m_Q \tilde{\xi}^\dagger + \tilde{\xi} m_Q \tilde{\xi}\right)$, 
with $\tilde{\xi} = \sqrt{\St}$, and the covariant derivative acts on $\cBt$ and $\cTt_\mu$ fields in the same manner
\begin{equation}
[\cDt_\mu \cBt (x)]^{ijk} = \partial_\mu \cBt^{ijk} (x) + i (B_\mu^i + B_\mu^j + B_\mu^k) \cBt^{ijk}(x)
.\end{equation}
The connection of the low-energy constants appearing in Eq.~\eqref{eqn:L} to those of $SU(2)$ \CPT\ is 
described in~\cite{Beane:2002vq}.

With twisted boundary conditions, 
the leading order \PQCPT\ interaction Lagrangian between the baryons and mesons now appears as
\begin{equation} \label{eq:Lint}
\cL 
= 
2 \a \left( \ol \cBt S^\mu \cBt \At_\mu \right)
+ 
2 \b \left( \ol \cBt S^\mu \At_\mu \cBt \right)
+ 
2 \cH \left( \ol \cTt_\nu S^\mu \At_\mu \cTt^\nu \right) 
+  
\sqrt{\frac{3}{2}} \cC 
\left[ 
\left( \ol \cTt_\nu \At^\nu \cBt \right)
+ 
\left( \ol \cBt \At^\nu \cTt_\nu \right)
\right]
,\end{equation}
where the twisted axial-vector meson field is defined by
$\At^\mu = \frac{i}{2} \left( \tilde{\xi} \Dt^\mu \tilde{\xi}^\dagger - \tilde{\xi}^\dagger \Dt^\mu \tilde{\xi} \right)$.
The familiar low-energy constants of $SU(2)$ can be identified as follows~\cite{Beane:2002vq}: 
$g_A = \frac{2}{3} \a - \frac{1}{3} \b$, $g_{\D N} = - \cC$, and $g_{\D \D} = \cH$. 
Notice there is an extra free parameter in the \PQCPT\ interaction Lagrangian compared to that of \CPT.

\section{Nucleon Axial Radius and Pseudoscalar Form Factor} \label{s:ax}

The axial current matrix elements of baryons have been studied extensively in chiral effective theories and many investigations
have now been tailored in an attempt to describe the numerical approximations in lattice QCD calculations, 
see \cite{Bijnens:1985kj,Jenkins:1991jv,Jenkins:1991es,Borasoy:1998pe,Kim:1998bz,Chen:2001yi,Beane:2002vq,Beane:2003xv,Beane:2004rf,Detmold:2004ap}. 
Determination of the nucleon axial charge has been the goal of numerous lattice investigations, for example
\cite{Woloshyn:1989ae,Martinelli:1988rr,Liu:1992ab,Fukugita:1994fh,Dong:1995rx,Gockeler:1995wg,Gusken:1999as,Horsley:2000pz,Dolgov:2002zm,Sasaki:2003jh,Khan:2004vw,Gurtler:2004ac}.  
The nucleon axial form factors have been calculated in \CPT\ \cite{Bernard:1994wn,Bernard:1998gv}, but have not been investigated on the lattice.
Here we consider the nucleon axial form factors in partially twisted \CPT. 
We detail how the axial radius and pseudoscalar form factor 
can be accessed, even at zero lattice momentum.

In partially quenched QCD, the iso-vector axial-vector current is defined by 
$J_5^{a \mu} = \ol Q \gamma^\mu \gamma_5 (\ol \tau {}^a / 2) Q$. The choice
of supermatrices $\ol \tau {}^a$ is not unique \cite{Golterman:2001qj}, 
even when one imposes the condition $\str \, \ol \tau {}^a = 0$.  
One should choose a form of the supermatrices that maintains the cancellation of valence and ghost quark loops
with an operator insertion~\cite{Tiburzi:2004mv}. For the flavor changing contributions we consider below, however, 
these operator self-contractions automatically vanish.  Thus regardless of the form chosen in extending $\ol \tau {}^a$, 
the only disconnected quark contributions are those from the gauge configurations. 
For our calculation we require the action of $J_5^{a  \mu}$ in only the valence sector, and specify
the upper $2 \times 2$ block of $\ol \tau {}^{a}$ to be the usual Pauli isospin matrices $\tau^a = (1, \bm{\tau})$. 
Henceforth we restrict our attention to the operator 
$J_5^{+ \mu} \equiv J_5^{1 \mu} + i J_5^{2 \mu} = \ol u \gamma^\mu \gamma_5 d$. 
The non-vanishing nucleon matrix element of this operator is the neutron to proton axial transition, 
which we decompose into two form factors
\begin{equation} \label{eq:decompose}
\langle p(P') | J_5^{+\mu} | n(P) \rangle 
= 
\ol u(P') 
\left[ 
2 S^\mu  G_A(q^2) + \frac{q^\mu \, q \cdot S}{(2 M_N)^2} G_P(q^2)
\right]
u(P)
,\end{equation} 
where $q^\mu = (P' - P)^\mu$ is the four-momentum transfer. Above $G_A(q^2)$ is the axial form
factor and $G_P(q^2)$ is the induced pseudoscalar form factor. 
The nucleon axial charge is $G_A(0)$, and in the chiral limit $G_A(0) = g_A$. 
Kinematically one can probe zero momentum transfer because
in the isospin limit the proton and neutron are degenerate on the lattice, as the electromagnetic mass difference 
is absent. 
We define the axial radius $<r_A^2>$ from the small momentum transfer expansion of the axial form factor
\begin{equation}
G_A(q^2) = G_A(0) + \frac{q^2}{6} <r_A^2> + \ldots \, \, 
.\end{equation}
Clearly to access the pseudoscalar form factor and axial radius, one needs momentum transfer
between the initial and final states.

In partially twisted \CPT, the form of the leading-order baryon axial current $J_5^{+\mu}$ is
\begin{eqnarray} 
J_5^{+\mu} 
&=&
2 \a \left( \ol \cBt S^\mu \cBt \, \ol \tau {}^+ \right) 
+ 
2 \b \left( \ol \cBt S^\mu \, \ol \tau {}^+ \cBt \right)
+ 
2 \cH \left( \ol \cTt_\nu S^\mu \, \ol \tau {}^+ \cTt^\nu \right) 
\notag \\
&& +  
\sqrt{\frac{3}{2}} \cC 
\left[ 
\left( \ol \cTt {}^\mu \, \ol \tau {}^+ \cBt \right)
+ 
\left( \ol \cBt \, \ol \tau {}^+ \cTt ^\mu \right)
\right] \label{eq:Bax}
,\end{eqnarray}
where the full chiral structure of the axial-vector current can be obtained under the replacement 
$\ol \tau {}^+ \to \frac{1}{2} \left( \tilde \xi ^\dagger \ol \tau {}^+ \tilde \xi  +  \tilde \xi \ol \tau {}^+ \tilde \xi^\dagger \right)$. 
To find contributions to the axial radius and pseudoscalar form factor, 
we must consider tree-level contributions of which there is the pion
pole term and local interaction terms. 
The latter are derived conveniently from the Lagrangian. Let $F^{\mu\nu}_A$ be the field-strength tensor 
of the external axial field.  In untwisted \CPT, one has the dimension-six interaction term contained in the
Lagrangian
\begin{equation}
\cL 
= 
\frac{2 n_A}{\L_\chi^2} 
\left( 
\ol N S_\mu \tau^+ N 
\right) 
\partial_\nu F^{\mu \nu}_A 
,\end{equation}
that leads to a contribution $\delta J_5^{+\mu}$ to the axial current at next-to-leading order
\begin{equation}
\delta J_5^{+\mu} 
=  
\frac{2 n_A}{\L_\chi^2} 
\left[ 
\partial^\mu \partial^\nu 
\left( 
\ol N S_\nu \tau^+ N 
\right) 
- 
\partial^2 
\left( 
\ol N S^\mu \tau^+ N 
\right) 
\right]
.\end{equation}
In partially twisted \CPT\, the analogous next-to-leading order axial current reads
\begin{eqnarray}
\delta J_5^{+\mu} 
&=&
\frac{1}{\L_\chi^2} 
\Bigg\{ 
2 n_\a 
\left[
\Dt^\mu \Dt^\nu 
\left( \ol \cBt S_\nu \cBt \, \ol \tau {}^+ \right)
-
\Dt^2
\left( \ol \cBt S^\mu \cBt \, \ol \tau {}^+  \right)
\right]
\notag \\
&& +
2 n_\b 
\left[
\Dt^\mu \Dt^\nu 
\left( \ol \cBt S_\nu \, \ol \tau {}^+ \cBt \right)
-
\Dt^2
\left( \ol \cBt S^\mu \, \ol \tau {}^+ \cBt \right)
\right]
\Bigg\}
.\end{eqnarray}
The relation of $n_A$ to the \PQCPT\ parameters can be found from matching, 
$n_A = \frac{2}{3} n_\a - \frac{1}{3} n_\b$.

Also, the terms of the axial current in Eq.~\eqref{eq:Bax} 
can be combined with those of the interaction Lagrangian in Eq.~\eqref{eq:Lint} 
to generate loop contributions to the axial radius and pseudoscalar form factor. 
These two observables, however, are rather special in heavy baryon \CPT, because
tree-level terms dominate over pion loop contributions. 
This is because the axial current is inserted on the baryon lines and generates 
terms $\sim q^2 / M_N^2$, rather than terms $\sim q^2 / m_\phi^2$ present in 
electromagnetic form factors, for example.  
Thus these quantities are largely insensitive to the long-range effects
introduced by twisting. Subsequently twisting can be used to produce momentum transfer between nucleon states without
sizable finite volume corrections. In calculating these quantities, we find
\begin{eqnarray}  \label{eq:answer}
<r_A^2> &=& \frac{6 n_A}{\L_\chi^2} \left[ 1 + \cO \left( \frac{m_\phi^2}{\L_\chi^2}, \frac{m_\phi^2}{M_N^2} \right) \right]  \\
G_P(q^2)&=& (2 M_N)^2 
\left(
\frac{g_A}{q^2 - m_\pi^2} - \frac{2 n_A}{\L_\chi^2} \right)
\left[ 1 + \cO \left( \frac{m_\phi^2}{\L_\chi^2}, \frac{q^2}{\L_\chi^2}, \frac{m_\phi^2}{M_N^2}, \frac{q^2}{M_N^2} \right) \right]
,\label{eq:answer2}\end{eqnarray}
where the recoil corrections ($\propto M_N^{-2}$) can be determined at the one-loop level~\cite{Borasoy:1998pe}, while
the equally sizable chiral corrections ($\propto \L_\chi^{-2}$) enter at two loops \cite{Bernard:1995dp}. Above $m_\phi$
is an abbreviation for the meson masses of \PQCPT. Because these \PQCPT\ results are determined at tree-level, sea quark masses
do not appear and  the dependence on the parameters $n_\a$ and $n_\b$ enters only through the combination
$\frac{2}{3} n_\a - \frac{1}{3} n_\b = n_A$. In this way, the partially quenched results maintain the Adler-Dothan relation for the 
pseudoscalar form factor~\cite{Adler:1966}.

In writing Eqs.~\eqref{eq:decompose}, \eqref{eq:answer}, and \eqref{eq:answer2}, 
we have used a compact notation for the momenta $P$ and $P'$, which have the form
\begin{eqnarray}
P^\mu  &=& \left( \sqrt{M_N^2 + \left(\frac{2 \pi \bm{n}}{L} + \bm{B}_\cN \right)^2}, \frac{2 \pi \bm{n}}{L} + \bm{B}_\cN \right) \\
P'^\mu &=& \left( \sqrt{M_N^2 + \left(\frac{2 \pi \bm{m}}{L} + \bm{B}_\cP \right)^2}, \frac{2 \pi \bm{m}}{L} + \bm{B}_\cP \right)
,\end{eqnarray}
as well as for the momentum transfer, which is  $q^\mu \approx \left( 0 , \bm{P'} - \bm{P} \right)$ in the non-relativistic limit.\footnote{%
Here we have neglected the $\bm{B}$ dependence of the proton and neutron masses that is dynamically generated from finite volume effects. 
These effects lead to $M_n - M_p \neq 0$ even in the isospin limit, but the difference introduced is beyond the order we are working. 
}
We have chosen the above notation to be consistent with Sec.~\ref{s:obs}. Finally let us spell out our results and the resulting form 
of Eq.~\eqref{eq:decompose}
at zero lattice momentum, i.e.~$\bm{m} = \bm{n} = \bm{0}$. To this end we define $B_{\pi^+}^\mu = (B_\cP - B_\cN)^\mu = (B^u - B^d)^\mu$, 
which is the induced $U(1)$ charge of the $\pi^+$ field due to twisting. At zero lattice momentum, we have
\begin{eqnarray} 
\langle p(\bm{0}) | J_5^{+\mu} | n(\bm{0}) \rangle 
&=& 
\ol u(\bm{0}) 
\Bigg\{ 
2 S^\mu  
\left[ G_A(0)
- \frac{1}{6} <r_A^2>  \bm{B}^2_{\pi^+}
\right] \notag \\
&&+ 
B_{\pi^+}^\mu \, \bm{B}_{\pi^+} \cdot \bm{S}  
\left[
\frac{g_A}{\bm{B}_{\pi^+}^2 + m_\pi^2}
+ 
\frac{1}{3} < r_A^2 >
\right]
\Bigg\}
u(\bm{0})
,\end{eqnarray}
where we have used $<r_A^2>$ as a replacement for the quantity in Eq.~\eqref{eq:answer}. 
This form makes transparent how varying the twisting parameters of the quarks can lead to a determination
of the axial radius and pseudoscalar form factor. We remark that these determinations suffer additional finite volume 
corrections but these do not occur until higher orders in the heavy baryon 
and chiral expansions. The simplicity of these results, moreover, can be used as a consistency check on the chiral
extrapolation. For example, if one sees valence or sea quark mass dependence of the axial radius greater than 
$\sim m_\phi^2 / M_N^2 \times 100\%$, then one is certainly not in the chiral regime.   
The same is true of the sea quark mass dependence of the pseudoscalar form factor, which has, however, 
dramatic behavior as a function of the valence quark masses.

\section{Summary} \label{summy}

For hadronic matrix elements of flavor changing operators, twisted valence quarks produce momentum transfer 
between the initial and final states. Thus such form factors' momentum transfer dependence can be explored without the additional generation 
of gauge configurations. 
Moreover the near forward limit can be explored to extract radii, \emph{etc}., without necessitating a box with a very long side
and subsequent momentum extrapolation. 
There are, however, modified finite volume corrections to these form factors arising 
from twisted boundary conditions.
For large enough volumes and below particle production thresholds, these corrections are exponentially small. 
We develop partially twisted baryon \CPT\ as a means to address the extrapolation of lattice QCD calculations
employing twisted valence quarks. 
We pursue two simple examples, the nucleon axial radius and pseudoscalar form factor. These quantities do not receive finite volume 
corrections until higher orders in the chiral expansion, thus providing a clean test case for the utilization 
of the induced momentum transfer from twisted quarks.  Moreover, the applicability of \CPT\ in the baryon sector 
can be tested due to the simple quark mass dependence predicted at leading order for these observables. 
Lastly confrontation with experimental data (which is ample for these observables)  
presents an essential test of lattice techniques.

\begin{acknowledgments}
We thank Shailesh Chandrasekharan and Tom Mehen for numerous discussions,
and Will Detmold for comments on the manuscript.   
This work is supported in part by the U.S.\ Dept.~of Energy,
Grant No.\ DE-FG02-96ER40945, and we acknowledge the Institute for 
Nuclear Theory at the University of Washington for its hospitality
during the completion of this work. 
\end{acknowledgments}

\appendix

\bibliography{hb}

\end{document}